\documentclass[prl,twocolumn,showpacs,amsmath,amssymb]{revtex4}

\usepackage{graphicx}
\usepackage{bm}

\begin{document}

\newcommand{\be}{\begin{equation}}
\newcommand{\bea}{\begin{eqnarray}}
\newcommand{\ee}{\end{equation}}
\newcommand{\eea}{\end{eqnarray}}

\title{Dynamical Motifs: Building Blocks of Complex Network Dynamics}

\author{Valentin~P.~Zhigulin}
\email{zhigulin@caltech.edu}
\affiliation{Department of Physics,
California Institute of Technology, Pasadena, CA 91125}
\affiliation{Institute for Nonlinear Science, University of
California, San Diego, La Jolla, CA  92093-0402}

\begin{abstract}

Spatio-temporal network dynamics is an emergent property of many complex systems which
remains poorly understood. We suggest a new approach to its study based on the analysis of 
dynamical motifs -- small subnetworks with periodic and chaotic dynamics. 
We simulate randomly connected neural networks and, with increasing density of connections, 
observe the transition from quiescence to periodic and chaotic dynamics. We explain this transition
by the appearance of dynamical motifs in the structure of these networks. 
We also observe domination of periodic dynamics in simulations of spatially distributed networks with local 
connectivity and explain it by absence of chaotic and presence of periodic motifs in their
structure.

\end{abstract}


\maketitle

Dynamics in networks underlie functioning of many complex systems such as the
brain~\cite{KochL99}, cellular regulatory machinery~\cite{Goldbeter02}, ecosystems~\cite{Kondoh03}
and many others. These systems exhibit a wide repertoire of 
dynamics, ranging from periodic oscillations in cell 
cycle and brain rhythms to chaos in food webs and chemical reactions.  Despite the recent
rapid advancements in our ability to elucidate statistical properties of the underlying networks
~\cite{AlbertB02,DorogovtsevM02,Newman03}, 
surprisingly little is understood about their dynamical behavior. This is due to several reasons,
in particular, inadequacy of the methods of nonequilibrium statistical mechanics in the 
domain of heterogeneous mesoscopic systems
and inability of the dynamical systems theory to deal with systems having more than order-1 dimensions. 

Fundamental problem which one faces while trying to understand dynamics in complex
networks is the strong influence of their structure on their non-Hamiltonian dynamics. This influence 
may induce long term connectivity-dependent spatio-temporal correlations which present
formidable problem for understanding of the dynamics. Statistical methods allow to solve this problem
in the limit of infinite-size networks~\cite{SompolinskyCS88}, but they are not applicable
to the study of realistic networks with non-uniform connectivity and a relatively small size.

It was recently found that many real networks include statistically significant subnetworks, 
so-called {\it motifs}, in their structure~\cite{MiloSIKCA02}.
In this Letter we suggest the use of {\it dynamical motifs} -- small subnetworks with
non-trivial dynamics -- as a new approach to the study
of recurrent dynamics in complex networks. In it we combine dynamical and statistical
methods to identify dynamical motifs and evaluate probability of their occurrence
in the structure of networks. We show that the emergence of periodic and 
chaotic dynamics in networks of increasing structural complexity is linked to the appearance
of periodic and chaotic motifs in their connectivity. We also consider spatially distributed
networks with local connectivity and show that chaotic motifs are absent in their structure.
We also suggest that this approach may be useful for study of the dynamics in networks of arbitrary
structure and size.

In many complex systems the dynamics of individual elements and the rules of their interaction
are relatively simple and the resulting complex behavior is an {\em emergent} consequence of these 
interactions. Hence, in order to study the influence of the structure on the dynamics of 
networks let us focus on models with simplest interactions and dynamics at each node. 
Let $x_i(t)\in [0;1]$, $i=1,\ldots,N$ be a set of variables describing properly scaled states 
of $N$ elements connected in a network. Consider
the time evolution of network's state vector $\bm{X}(t)=\{x_1(t),x_2(t),\ldots,x_N(t)\}$
described by the following set of first order differential equations
\be
\frac{d\bm{X}(t)}{dt} = - \bm{X}(t) + \bm{F}(\bm{X}(t)),
\label{eq1}
\ee
where $\bm{F}(\bm{X})=\{f_1(\bm{X}),f_2(\bm{X}),\ldots,f_N(\bm{X})\}$ is a set of sigmoid
nonlinearities with $[0;1]$ value ranges. This general class of models includes 
continuous version of Random Boolean (Genetic) Networks (cRBN)~\cite{MestlBG97,GlassH98}, in which 
$f_i(\bm{X})$ are randomly chosen Boolean functions of their arguments, and continuous-time 
Artificial Neural Networks (cANN)~\cite{Hopfield84}, in which  
$f_i(\bm{X})=f(({\Hat{W} \cdot \bm{X}})_i + \sigma_i)$, where $\Hat{W}$ is the coupling matrix and
$\sigma_i$ are thresholds. Both of these models were shown to exhibit complex periodic and 
chaotic dynamics in the biologically relevant cases of intermediate 
probabilities of gene expression in cRBN~\cite{GlassH98} and non-symmetric interactions 
in cANN~\cite{SompolinskyCS88}. 

To illustrate the use of dynamical motifs we employ a
simple cANN model with $f(x)=(1+\exp (-20\,x))^{-1}$, uniform external excitation 
$\sigma_i=\sigma=0.5$ and inhibitory interactions of the same 
strength: $\Hat{W}= - \Hat{G} w$, where $w=5$ and $\Hat{G}$ is the adjacency matrix
of the directed graph on which the network is defined. In this setting the model is similar to
the simplified version of a balanced network model~\cite{vanVreeswijkS96} with excitatory connections
replaced by a uniform field and can be viewed as a simple model of a cortical microcircuit. It is also an
extension of the concept of winnerless competitive networks~\cite{RabinovichVL01} to the case of arbitrary 
connectivity.  However, methods presented in this Letter can be used for other models as well.

\begin{figure}
  \begin{center}
\leavevmode
    \includegraphics{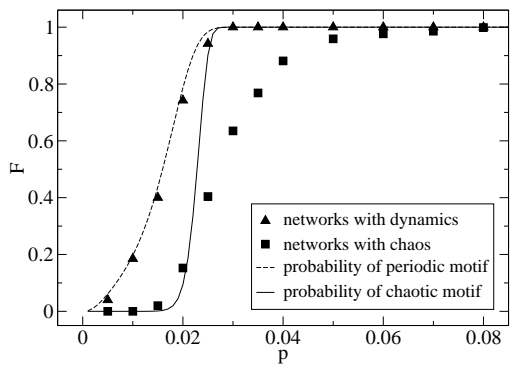}
  \end{center}
\caption{\label{LC}Fraction of networks $F$ with either periodic or chaotic ($\blacktriangle$) 
and only chaotic ($\blacksquare$) dynamics in simulations of 200-node cANN networks with node-to-node
connection probability $p$. Lines represent predictions based on the probability of
occurrence of periodic (dashed line, Eq.~(\ref{P_33})) and chaotic (solid line, Eq.~(\ref{P_811})) motifs.}
\end{figure}

We have performed Monte Carlo simulations of the described above cANN model defined on an ensemble
of random networks with $N=200$ nodes and uniform probability $p$ of node-to-node connections, {\it i.e.},
an Erd\H{o}s-R\'{e}nyi (ER) ensemble.
A sample of $2\cdot 10^4p$ random networks was generated for each considered $p$ and cANN dynamics
was simulated 100 times on
each of the networks, each time starting with different initial condition taken at random from 
the hypercube $R_{(0;1)}^{200}$. Sets of initial conditions were considered in order to eliminate the
influence of the basins of attraction in multistable networks which in itself is a complicated issue
requiring separate research. Largest Lyapunov exponent $\lambda$ was calculated in each simulation.
Networks with at least one initial condition leading
to $\lambda \in (-0.005;0.005)$, typically $\lambda \sim 10^{-4}$ were classified as having 
limit cycle dynamics and with $\lambda > 0.005$, typically $\lambda \sim  10^{-1}$ as having chaotic dynamics.
As the probability of connections $p$ was increased, the transition from fixed point to periodic and chaotic
dynamics was observed around $p=0.015$ (Fig.~\ref{LC}, triangles). The transition to solely chaotic dynamics
occurred around $p=0.025$ (Fig.~\ref{LC}, squares). 
%%%%%%%%%%%%%%%%%%%%%%%%%%%%%%%%%%%%%%%%%%%%%%%%%
%%%%%%%%%%%%%%%%%%%%%%%%%%%%%%%%%%%%%%%%%%%%%%%%%%

We now apply the concept of dynamical motifs in order to explain such
observations and make further predictions about the dynamics in networks. 
The main idea behind this approach is that the transition to periodic or chaotic
overall dynamics in a network occurs due to the appearance in its structure of small, not necessary isolated,
subnetworks which have the same type of dynamics. We call these subnetworks {\em dynamical motifs}.
Of course many dynamical motifs may be present in a given 
dynamical network, but at least one is needed in order for the network to have a given type of dynamics.
The dynamical phase transitions that are observed in models of complex networks are then
identified with the percolations of dynamical motifs, {\it i.e.,} $0$-to-$1$ transitions in the 
probability of their occurrence.
%%%%%%%%%%%%%%%%%%%%%%%%%%%%%%%%%%%%%%%%%%%%%%%%%
\begin{figure}
  \begin{center}
\leavevmode
    \includegraphics{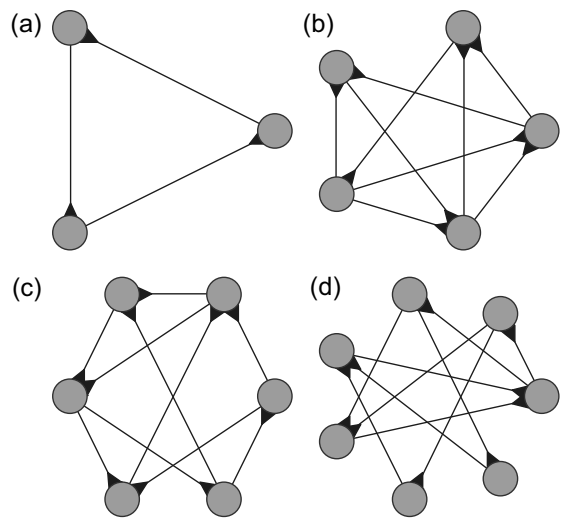}
  \end{center}
\caption{\label{fig2}Smallest dynamical motifs with periodic (a) and chaotic (b-d) dynamics. Triangles
indicate direction of the inhibitory connections.}
\end{figure}
%%%%%%%%%%%%%%%%%%%%%%%%%%%%%%%%%%%%%%%%%%%%%%%%%%

Let us consider an ER network with the probability of node-to-node connection $p$. The probability
$p_n$ that some $n$ nodes of this network form a given motif with $l$ links and no self-loops is equal to 
$p^l(1-p)^{n(n-1)-l}$. It is possible that some of the nodes in the motif are suppressed by connections
from a number $f$ of nodes outside of a motif that are frozen in the $"+1"$ state. Such connections 
would suppress dynamics 
of the motif and should be ruled out. By excluding them we obtain corrected probability
\be 
p_{nf}=p^l(1-p)^{n(n+f-1)-l}.
\label{p_n}
\ee
It is difficult to calculate exactly the probability to find $N_{(n;l)}$ such motifs in a network.
However, approximate calculation of the probability $P_{(n;l)}$ to
encounter one or more motifs is straight forward and 
is expected to work well in the case of sparsely connected networks ($p<<1$):
\bea \label{P_m}
P_{(n;l)} & = & 
1 - p(N_{(n;l)} = 0) \approx 1 - (1-p_{nf})^{\frac{N!}{A(N-n)!}} \nonumber \\
&\approx& 1 - \exp \left(-\frac{p_{nf}N!}{A(N-n)!}\right),
\eea
where $N!/(A(N-n)!)$ is the number of ways to pick $n$ nodes of the motif from the $N$-node
network multiplied by $n!/A$ -- number of ways to label a motif, with $A$ being the order of 
motif's automorphism group. This formula predicts percolation of these motifs 
at some intermediate value of $p$ which depends on $n$ and $l$.
We are interested in the values of $n$ and $l$ for which the percolation occurs at the smallest $p=p_c$.
Let us define it by the point where $P_{(n;l)}=1/2$. Then 
\be
\frac{N!}{(N-n)!\,A}p_c^l(1-p_c)^{n(n+f-1)-l}=\log 2.
\label{p_c}
\ee
Assuming $N \gg n$ and $p_c \ll 1$ we find that $p_c \sim A^{1/l}N^{-n/l}$. To minimize $p_c$, $n/l$
should be maximized and hence subnetworks with most nodes and least links 
will appear first as $p$ is increased from $0$.

Subnetworks with limit cycle dynamics include 3-loop (Fig.~\ref{fig2}(a)), 4-loop and other, more
complicated, structures. Since for the $n$-loop $l=n$ and $A=n$, 3-loop has smallest $p_c$.
According to (\ref{P_m}), its probability of occurrence is given by
\be
P_{(3;3)} \approx 1 - \exp \left(-\frac{N^3}{3}p^3(1-p)^{3(1+f)}\right)
\label{P_33}
\ee
and is plotted in Fig.~\ref{LC} by a dashed line (with $f$ evaluated from simulations).
This estimate predicts the appearance of dynamics in ER networks very accurately.

In order to find chaotic motifs we used the {\em nauty} package~\cite{McKay98} to generate
all possible non-isomorphic directed graphs with up to 8 nodes and 11 links,
simulated the cANN dynamics on them and calculated the largest Lyapunov exponent $\lambda$ for each. Digraphs
with $\lambda>0.005$, typically $\lambda \sim 10^{-1}$ were classified as chaotic.
In Fig.~\ref{fig2}(b-d) we show the first three chaotic motifs in the order of increasing number of 
nodes (5, 6 and 7 nodes) and minimal number of links (9, 10 and 10 links). 
Also, we found six non-isomorphic chaotic motifs with 8 nodes and 11 links, four of them with $A=2$ and 
two with $A=1$. Numerical evaluations of $p_c$ 
according to (\ref{p_c}) indicate that the latter motifs have smallest $p_c \approx 0.025$. Probability to 
find one or more such motifs in an ER network is given by 
\be
P_{(8;11)} \approx 1 - \exp \left(-4 N^8 p^{11}(1-p)^{45+8f}\right)
\label{P_811}
\ee
and is piloted in Fig.~\ref{LC} by a solid line (with $f$ evaluated from simulations of chaotic 
networks). As can be seen from the plot, this prediction is reasonably good. The discrepancy may be
caused by the approximate nature of the estimate (\ref{P_m}), severe undersampling
of the network space in simulations and disregard of the fact that not only frozen, but also
periodically oscillating external nodes suppress chaos in these motifs. 

\begin{figure}
  \begin{center}
\leavevmode
    \includegraphics{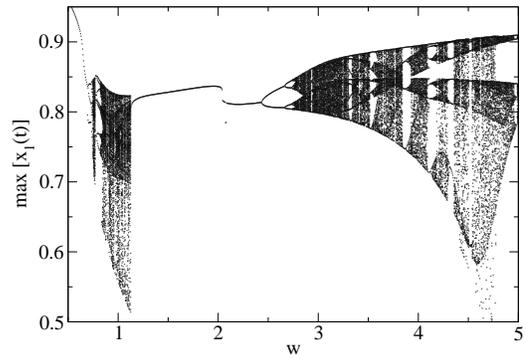}
  \end{center}
\caption{\label{fig3}Bifurcation diagram depicting local maxima of $x_1(t)$ oscillations
for a range of strengths $w$ of the inhibitory connections in the first chaotic motif (Fig.~\ref{LC}(b)).
Period doubling cascades start around $w=0.7$ and $w=2.5$.}
\end{figure}

While the origin of periodic dynamics in $n$-loops is obvious as they merely are negative
feedback loops, the nature of chaos in chaotic motifs is not self-evident. 
We have traced the route to chaos
in the first chaotic motif by lowering $w$ to $0.5$ and then gradually increasing it. In Fig.~\ref{fig3} 
the bifurcation diagram of the local maxima of $x_1(t)$ oscillations is plotted. This diagram reveals
two period doubling cascades, one starting around $w=0.7$ and another one around $w=2.5$.
%%%%%%%%%%%%%%%%%%%%%%%%%%%%%%%%%%%%%%%%%%%%%%%%%
%%%%%%%%%%%%%%%%%%%%%%%%%%%%%%%%%%%%%%%%%%%%%%%%%%

However, there are no conclusive experimental observations of chaotic
dynamics in either genetic or neuronal networks. Presence of chaotic 
dynamics would be inconsistent with the requirements of robustness and reproducibility of the 
response imposed on the living organisms by the environment. On the other hand, periodic dynamics
in these networks are very common. Hence, some of the assumptions that were used in these simplified
models must be wrong. As we show below, one of them is the assumption of uniformly random connectivity.
Recent experimental data suggests that 
metabolic networks possess scale-free structure~\cite{JeongTAOB00} while 
neuronal networks are highly clustered on both small~\cite{ShefiGSBA02} and large~\cite{HilgetagBO00}
spatial scales. Moreover, neurons in the brain
frequently form ordered spatial structures with distance-dependent probabilities of connections, 
so-called cortical microcircuits~\cite{SilberbergGM02,MaassNM02}. 

To illustrate the influence of spatial structure on the dynamical
properties of networks we simulated the cANN model on the 12-by-12 two-dimensional square lattice
with neuron-to-neuron connection probabilities obeying Gaussian distribution and forbidden 
self-connections:
\be
p(d_{ij}) = K N (1-\delta_{ij}) e^{-\left(\frac{d_{ij}}{\gamma}\right)^2}/\sum_{m,n=1}^{N} 
(1-\delta_{mn}) e^{-\left(\frac{d_{mn}}{\gamma}\right)^2}
\label{p_d}
\ee
where $d_{ij}$ is a metric distance between neurons $i$ and $j$, $\gamma$ is the length scale of
the distribution, $K$ is the average number of connections per neuron and $\delta_{ij}$ is the
Kronecker delta. In effect, $\gamma$
controls clustering of the connectivity, with values close to $1$ corresponding to networks
with mostly local connectivity and large values effectively diminishing the role of spatial 
structure and corresponding to ER-like connectivity. 

We generated a random sample of 1000 such networks with $\gamma=2$ and $K=14$. 
As in the case of an ER sample, cANN dynamics was simulated in each of the networks for 100
different random initial conditions and the largest Lyapunov
exponent was calculated in each simulation. An average connectivity $K=14$ corresponds
to $p\approx 0.1$ which has led to approximately $99\%$ of chaotic networks in a sample of ER
ensemble (Fig.~\ref{LC}(b)). On the contrary, around $99\%$ of the networks with $\gamma=2$
exhibited periodic dynamics and only about $1\%$
were chaotic. This result indicates that clustering plays an important role in defining dynamical
properties of neural networks. It may support an observation that many real neuronal networks
are locally clustered and exhibit reproducible, except for stochastic effects, dynamics.   

We now show how the idea of dynamical motifs may be used to understand these observations. 
Let us enumerate grid nodes by $1,\ldots,N$ and consider a motif with $n$ nodes and $l$ 
links placed on the grid.  We label its nodes by $i_1,i_2,\ldots,i_n$ and
its links by $i_{a_1}i_{b_1}$, $i_{a_2}i_{b_2}$, \ldots, $i_{a_l}i_{b_l}$ with
$a_j b_j$, $j=i\ldots l$ denoting ordered pairs of nodes that are connected.
Probability $P_{i_a i_b}$ for the grid nodes $i_a$ and $i_b$ to be connected is 
distance-dependent and is given by (\ref{p_d}). Then the
average probability that $n$ nodes of a network form this motif is obtained by averaging
over all possible placements of its nodes on a grid:
\be
p_n \approx \frac{1}{N^n} \sum_{i_1,\ldots,i_n=1}^N \left(\prod_{j=1}^{l} P_{i_{a_j}i_{b_j}}
\prod_{k=1}^{s} (1-P_{i_{u_k}i_{v_k}})\right),
\label{p_n2}
\ee
where $s=n(n-1)-l$ and $u_k v_k$ are all the pairs of unconnected nodes.
For example, an average probability for a 3-loop (Fig.~\ref{fig2}(a)) is
$
p_3 = \frac{1}{N^3} \sum_{i_1,i_2,i_3=1}^N P_{i_1i_2}P_{i_2i_3}P_{i_3i_1}
(1-P_{i_2i_1})(1-P_{i_3i_2})(1-P_{i_1i_3}).
$
As expected, in the limit of distance-independent probabilities expression (\ref{p_n2})
becomes equivalent to the ER case.
Probability of occurrence of one or more motifs in a network can again be approximated by
(\ref{P_m}). It is approximately 1 for 3- and 4-loops and $P_m\approx 0$ for the (5; 9) 
motif in networks with $\lambda=2$. Other chaotic motifs would be even less probable
because connection probability falls off quickly with distance in these networks.
Also, $P_m\approx 1$ for the (5; 9) motif in networks with $\lambda=12$. Hence, periodic motifs are 
present and chaotic motifs are absent in spatial networks with local connectivity ($\lambda=2$), but
chaotic motifs are present in non-local networks with $\lambda=12$. 
These calculations may also explain recent observations of periodic and chaotic dynamics in 
models of cortical neural microcircuits with local and non-local spatial organization of connectivity~\cite{MaassNM02}. Eq.~(\ref{p_n2}) may also be applied in the study of networks with other
distributions of connectivity.

In conclusion, a method to study dynamical behavior of networks
by examining minimal building blocks of the dynamics was suggested. 
Calculations of abundance of dynamical motifs in networks with different structure allow
to study and control dynamics in these networks by choosing connectivity that maximizes
the probability of motifs with desirable dynamics and
minimizes probability of motifs with unacceptable dynamics. 
Using this method we predict that connectivity of cortical microcircuits might be such that it minimizes 
the occurrence of chaotic motifs in their structure.

It was shown in~\cite{MiloSIKCA02} that many real networks have 3-loops in their structure.
In most cases there are very few of such loops and it was argued that their presence
is not statistically significant. However, we suggest that such dynamical motifs are important
because presence of even one or two of them may profoundly
influence dynamical behavior of the whole network by slaving dynamics of many adjacent nodes.

\begin{acknowledgments}
The author would like to thank Ramon Huerta, Mikhail Rabinovich, Michael Cross and Dmitri Chklovskii 
for usefull discussions, Brendan McKay for extending his {\it nauty} package with digraphs and 
Gilles Laurent for continuous inspiration and support. This work was partially supported 
by NSF grants EIA-0130708 and EIA-0130746 and DOE grant DE-FG03-96ER14592.
\end{acknowledgments}

\end{document}